\documentclass[11pt]{article}
\usepackage{amsmath,amsfonts,amssymb,eepic,graphicx,epsfig,epic,hyperref}
\newcommand{\sepAuthor}{0.3in}
\newcommand{\sepAbstract}{0.3in}
\newcommand{\skipKeywords}{20pt}

\long\def\mytitlepage#1#2#3#4{
        \thispagestyle{empty}
        \begin{center}
        {\Large\bf #1}

        \vspace{\sepAuthor}
        #2\\
        \medskip

        \vspace{\sepAbstract}
        {\Large Abstract}
        \end{center}

        \noindent{#3}
        \vskip\skipKeywords

        \noindent{#4}
        \clearpage
        }
\usepackage{amsthm}
\theoremstyle{plain}
\newtheorem{theorem}{Theorem}[section]
\newtheorem{lemma}[theorem]{Lemma}
\newtheorem{corollary}[theorem]{Corollary}
\newtheorem{proposition}[theorem]{Proposition}

\theoremstyle{definition}
\newtheorem{definition}[theorem]{Definition}

\theoremstyle{remark}
\newtheorem{remark}[theorem]{Remark}

\def\squareforqed{\hbox{\rlap{$\sqcap$}$\sqcup$}}
\def\qed{\ifmmode\squareforqed\else{\unskip\nobreak\hfil
\penalty50\hskip1em\null\nobreak\hfil\squareforqed
\parfillskip=0pt\finalhyphendemerits=0\endgraf}\fi}

\newenvironment{proofof}[1]{\begin{trivlist}%
\item[]{\flushleft\em Proof of #1. }}
{\end{trivlist}}
\newcommand{\comments}[1]{}

\newcommand{\Tr}{\textrm{Tr}}

\newcommand{\Prob}{\mathrm{Prob}}
\newcommand{\eps}{\epsilon}

\newcommand{\ccc}{\mathrm{Com}}

\newcommand{\trace}{\mathrm{tr}}
\newcommand{\tr}{\mathrm{tr}}
\newcommand{\linear}{\mathbf{L}}

\newcommand{\Space}[1]{\mathcal{#1}}

\newcommand{\superNorm}[1]{\|#1\|_\diamond}

\newcommand{\defeq}{\stackrel{\mathrm{def}}{=}}
\newcommand{\sign}{\mathrm{sign}}
\begin{document}
\mytitlepage{Tensor Norms and the Classical Communication Complexity
of Nonlocal Quantum Measurement\footnote{A preliminary version of this paper appeared as part
of an article in {\em Proceedings of the
the 37th ACM Symposium on Theory of Computing (STOC 2005)}, 460--467, 2005.}}{
Yaoyun Shi \qquad\ and\qquad Yufan Zhu\\
       {Department of Electrical Engineering and Computer Science}\\
       {University of Michigan}\\
       {1301 Beal Avenue}\\
       {Ann Arbor, MI 48109-2122, USA}\\
       {Email: \{shiyy$|$yufanzhu\}@eecs.umich.edu}
}{We initiate the study of quantifying nonlocalness
of a bipartite measurement by the minimum 
amount of classical communication
required to simulate the measurement.
We derive general upper bounds, which are expressed
in terms of certain {\em tensor norms} of the measurement
operator. As applications, we show that (a) If the amount of
communication is constant, quantum and
classical communication protocols with unlimited amount
of shared entanglement or shared randomness compute the same
set of functions; (b) 
A local hidden variable model needs only a constant amount of communication
to create, within an arbitrarily small statistical distance, a distribution
resulted from local measurements of an entangled quantum state, as long as
the number of measurement outcomes is constant.

}{{\bf Keywords:} Quantum entanglement, classical simulation, 
communication complexity, tensor norms, Bell Inequality}

\section{Introduction and summary of results}
\begin{trivlist}
\item{\bf Background.}
Although Einstein himself made significant contributions
to the development of quantum mechanics, he famously questioned
the ``completeness'' of the theory with
a ``paradox'' that he formulated with Podolsky and Rosen
\cite{EPR35a}. Following Bohm \cite{Bohm51},
the essence of the paradox is:
two ``quantum coins'', possessed 
by two parties Alice and Bob, 
may be correlated in a state
that can be schematically represented as 
\[\frac{1}{\sqrt{2}}\ \left(|\texttt{Head}\rangle_A|\texttt{Tail}\rangle_B
- |\texttt{Tail}\rangle_A|\texttt{Head}\rangle_B\right).\]
If each party measures his or her
coin, with $1/2$ probability, one of the two outcomes would be observed.
However, once a measurement is made
by one party, say, Alice, then Bob would
always observe the opposite outcome with certainty.
A unique property of the state is that, no matter
what property of the coins is measured -- be it determining
their positions or the velocities -- Bob's outcome is also opposite to
that of Alice {\em with certainty}.
Since what Alice does locally should not affect Bob's world,
this is at odds with the ``uncertainty'' principle of quantum
mechanics that not all pairs of properties can be determined with
certainty.

The Einstein-Podolsky-Rosen (EPR) paradox did not 
reduce quantum mechanics to contradictions. Instead, it revealed
the essence --- quantum entanglement --- 
that underlies the many counter-intuitive properties and marvelous 
capabilities of quantum information. 
For example, in his far reaching paper \cite{Bel64a}, John Bell formulated
a set of inequalities, referred to as {\em Bell Inequalities} now,
that must be satisfied by the correlations produced by any so called
{\em hidden variable} classical model 
but would nevertheless be violated by
some quantum correlations. The latter has been
confirmed by several experiments (e.g., \cite{Tittel98}).
Another seminal example is
the {\em quantum key distribution} protocol \cite{BB84a}, which
has been shown to be
{\em information theoretically secure} \cite{LC99a, Mayers01}, 
as a consequence of properties of quantum entanglement.

Given its importance, quantum entanglement has been the subject
of numerous studies (see, e.g., the books \cite{NC00a, Prec}).
The focus has been on understanding the inherent quantitative
tradeoffs among various resources involved in the creation and conversion
of entangled states.
As entanglement is the result of nonlocal quantum interactions,
understanding various aspects of 
the nonlocality of quantum operations is also
of fundamental importance.
Quantifying nonlocality of quantum operators is precisely
the purpose of this paper.

A natural nonlocality measure of a quantum operation
is its {\em generating capacity}, which is the maximum entanglement 
increase that it could create (see e.g., \cite{BHLS02a}).
Another approach, more from a computational point of view,
is to consider the amount of resources,
such as the time in the case of using elementary
Hamiltonians, or the number of elementary gates,
required to simulate the operator (e.g., \cite{CHN03, CLVV03}).

\item{\bf Main result.}
In this paper, we take a completely different approach
to quantify the nonlocality of quantum operations,
following intuitions from the subject of communication complexity.
Our work is not the first to apply
communication complexity to the study of entanglement.
There has been a line of research, which we will review
shortly, that studies the classical communication
complexity of simulating quantum correlations.
Nevertheless, our emphasis is on quantum operators,
and we focus on measurement operators, 
while our approach can be extended to the most general quantum operations.

Consider the following quantum process.
Alice and Bob share a bipartite state $|E\rangle_{AB}$.
They apply local operations $R_A$ and $R_B$ to his/her system, 
before a final measurement $Q$ is applied to the joint system,
producing a distribution $\mu=\mu(Q, |E\rangle, R_A, R_B)$ 
of measurement outcomes.

Imagine now that Alice and Bob loose their quantum power.
They both know classical descriptions of $Q$ and $|E\rangle$,
and that of their local operations, but do not know
what the other party's local operation is.
From those classical information, they hope to simulate the 
quantum process, by producing an output whose distribution
is close to $\mu$, through a communication process
that starts with an unlimited supply of common random bits.
We define the {\em classical communication complexity} 
of $Q$, denoted by $\ccc(Q)$, to be the minimum number
of bits that need to be exchanged by the simulating 
communication process. 

Intuitively, $\ccc(Q)$ reflects how nonlocal
$Q$ is.  Consider, for example, the simple case
that $Q$ consists of local operations.
If there is no quantum correlation in the initial
state, it is clear that Alice and
Bob could simulate the quantum process without interaction.
We shall see that even if the initial state is entangled,
they do need only exchange a constant number of bits.

On the other hand, $\ccc(Q)$ could be much larger.
Let $n\ge1$ be an integer. Consider the following operator.
\begin{equation}\label{eq:ipn}
\textsc{IP}_n \defeq \sum_{x, y\in \{0, 1\}^n, x\cdot y=1}
\ |x\rangle\langle x| \otimes |y\rangle\langle y|.
\end{equation}
When $R_A$ is to create a state $|x\rangle$, $x\in\{0, 1\}^n$,
and $R_B$ creates $|y\rangle$, $y\in\{0, 1\}^n$,
then $\textsc(IP)_n$ determines if $x\cdot y=1$. This is the
so called ``Inner Product'' function well studied
in the communication complexity literature.
It is well known that any classical communication
protocol for solving Inner Product requires $\Omega(n)$
bits of communication. In fact, Cleve, van Dam,
Nielsen, and Tapp \cite{CvDNT99a} proved that
$\Omega(n)$ quantum bits are necessary, too.
Thus $\ccc(\textsc{IP})=\Omega(n)$.
We do not know if this bound for $\ccc(\textsc{IP})$ is tight.

The goal of this paper is to understand how $\ccc(Q)$ 
is determined in general.  It is not immediately clear
if $\ccc(Q)$ can be bounded from above for all $Q$,
as the dimension of the initial state $|E\rangle$
 could be arbitrarily
large.
Our main result is to derive a general upper
bound on $\ccc(Q)$ in terms of a certain operator norm $\|Q\|_\diamond$
on $Q$, which is bounded from above polynomially in $Q$'s dimension.

\begin{theorem}[Informally]\label{thm:diamondInf}
\newcommand{\sd}{\mathbf{SD}}
For any bipartite quantum measurement $Q$,
$\ccc(Q)=O(\|Q\|^2_\diamond)$.
In particular,
if $K$ is the dimension of the space that $Q$ acts on,
$\ccc(Q)=O(K^4)$.
\end{theorem}

The diamond norm $\|Q\|_\diamond$ is originally defined on superoperators,
and has been a powerful tool in the study
of quantum interactive proof systems \cite{Kitaev:2000:PAE} and
quantum circuits on mixed states \cite{AKN98a}.
We make use a natural mapping from bipartite operators to
superoperators to define norms on the former
based on norms on the latter.

The approach in proving Theorem~\ref{thm:diamondInf}
can be extended to obtain general upper bounds
on $\ccc(Q)$ in terms of other operators norms.
Those norms belong to so called {\em tensor norms}, i.e.,
norms $\|\cdot\|_\alpha$ that satisfies
$\|P\|_\alpha\le\|A\|\cdot\|B\|$, whenever
$P=A\otimes B$. Tensor norms have been studied for decades
with a great deal of rich concepts and deep results (see, e.g.,
\cite{Defant:1993}). In recent years,
they have been applied to quantum information theory to 
characterize and quantify the nonlocality of quantum states 
\cite{Rud00a, Rud02a}.
The tensor norms that appear in our upper bounds
capture the nonlocality
of bipartite operators in their own way, 
and may have further applications.

\item{\bf Applications on quantum communication complexity.}
After obtaining those general upper bounds,
we show that they in turn have
useful applications
on quantum communication complexity.
Recall that in the setting
of communication complexity \cite{Yao:1979:comm, Yao:1993:circuit},
Alice and Bob wish to
compute a function $f(x, y)$, where $x$ is known to Alice only,
and $y$ is known only to Bob.
The communication complexity of $f$ is the minimum
amount of information that Alice and Bob need to exchange
in order to compute $f$ correctly for any input.
Communication complexity has been a 
major research field (see, e.g., the book \cite{kushilevitz:1997:book}),
with many problems of rich structures and deep connections
to other aspects of complexity theory. 

A concrete application of our result is on the advantage of sharing entanglement 
in quantum protocols, a question that has
puzzled many researchers \cite{CB97a,BCvD01a, Klauck01, NayakS02}.
It is known that sharing entanglement could give a constant
additive advantage \cite{CB97a,BCvD01a}, 
or save a half of the communication \cite{CvDNT99a}.
However, little is known on the limit of the advantage.
This is in sharp contrast with the classical case
of sharing randomness, where we know that it can
only save at most a logarithmic additive term 
\cite{Newman:1991}.
If there is a quantum protocol that exchanges
$q$ qubits with $m$  qubits of prior entanglement, then
the best classical simulation we know is $\exp(\Omega(q+m))$.
This is embarrassingly large, especially when $q<<m$.
Using our upper bound on
the classical communication complexity of nonlocal operators,
we prove the following result. 
Note that in the 
{\em Simultaneous Message Passing (SMP)} model with shared randomness,
the two parties holding the inputs share an arbitrarily
long random string, and each send a single message to
a third party,
who is required to determine the outcome correctly with 
high probability.
\begin{theorem}
\label{thm:twoway}
If a twoway
quantum protocol uses $q$ qubits of communication and  $m$ qubits of
share entanglement, then it can be simulated by a classical protocol using
$\exp(O(q))$ bits with shared randomness. 
The simulation does {\em not} depend on $m$.
Furthermore, it can be carried out in the SMP model with
with shared randomness.
\end{theorem}
Notice that the exponential dependence on $q$ can not be
improved, because of the existence of an
exponential separation of quantum
and classical communication complexities
for some partial function,
discovered by Raz \cite{Raz:1999:com}.  As a consequence of the above theorem,
\begin{corollary}
\label{co:constLog}
If a communication complexity problem has a constant
cost quantum communication protocol with
shared entanglement, it also has a constant cost classical
protocol with shared randomness.
\end{corollary}

It is interesting to contrast the above with a recent
result by Yao \cite{Yao:2003:finger}, which is of 
a similar type but of the opposite direction. 
\begin{theorem}[\cite{Yao:2003:finger}]
\label{thm:YaoFinger}
If a communication complexity problem of input size $n$
has a constant cost 
classical SMP protocol with shared randomness, it
has an $O(\log n)$ cost quantum SMP protocol
{\em without} shared entanglement.
\end{theorem}
Combining this result with ours, we have
\begin{corollary}
\label{res:q2qsmp}
If a communication complexity problem of input size $n$
has a constant cost
twoway quantum protocol with shared entanglement, it has an
$O(\log n)$ cost quantum SMP protocol without shared entanglement.
\end{corollary}

\item{\bf Applications on simulating quantum correlations}.
Yet another application of our classical simulation
of quantum measurements is to give efficient simulations
of quantum correlations by the hidden variable model
assisted with classical communication.
The scenario is as follows. Suppose Alice and Bob
are given an entangled quantum state. Then each of them,
without any communication, 
applies to their portion of the state some local measurement not known
to the other party. The result is a correlated joint
distribution on both measurement outcomes.
There are such correlations that violate the Bell Inequalities, hence
impossible to  generate by
any reasonable classical procedure in which Alice and Bob do 
not communicate.

A natural next step to extend the above work of Bell
is to investigate the minimum amount of classical communication
required to simulate a quantum correlation. Most of 
the works addressing this question 
focus on the exact simulation and on measuring a constant number
of qubits~\cite{Toner03, Bacon03, Csirik02, Steiner:1999hg, 
BCT99a, Maudlin92}. We study the approximate and asymptotic simulation
of quantum correlations,  where the joint random variables
take a constant number of possible values but are nevertheless
produced from (the two party) sharing an entangled state of
an arbitrary dimension and applying arbitrary local measurements.

\begin{theorem}[Informally]
\label{thm:bell}
In the above scenario, a  
$O\left(\ln\frac{1}{\eps}/\eps^2\right)$  number of classical bits is sufficient 
to approximate the quantum correlation with a $\eps$ statistical
distance.
\end{theorem}

\item{\bf Organization.}
The rest of the paper is organized as follows. We start with the description
of a general framework for classical simulation of quantum protocols.
The cost parameter of this framework is then optimized in the next section,
giving the main theorem. In the section that follows we give applications of the
theorem. Finally we conclude with several open problems.
\end{trivlist}

\section{A simulation framework}
\label{sec:simulate}
Our classical simulation of quantum protocols falls into the following
framework.  Let $p$ be the acceptance
probability (i.e., the probability of outputting $1$)
of a given quantum protocol (which arises either from 
a communication task or from a bipartite measurement). 
We express $p=\langle \psi_A|\psi_B\rangle$, for two vectors $|\psi_A\rangle$ and $|\psi_B\rangle$ that can
be prepared by Alice and Bob by herself/himself. Note that the lengths
of the two vectors may be very large, in general. Indeed the shorter
their lengths are, the better our simulation is.

More precisely, if for some number $C$, 
$\||\psi_A\rangle\|\le C$ and $\||\psi_B\rangle\|\le C$, then
the following simulation uses $O(C^4)$ bits.
 Alice and Bob send Charlie $\||\psi_A\rangle\|$ and $\||\psi_B\rangle\|$,
respectively, up to $O(1/C)$ precision. This requires $O(\log C)$ bits.
They then proceed to estimate $\cos\theta$, for 
the angle $\theta$ between $|\psi_A\rangle$ and
$|\psi_B\rangle$ up to a precision of $O(1/C^2)$.
The protocol in Kremer, Nisan and Ron\cite{Kremer:1995:comm}, 
which is based on 
the following observation of 
Goemans and Williamson \cite{GoemansW94}, gives a protocol
that accomplishes the latter task using $O(C^4)$ bits.

Assume for simplicity that all vectors are real
(the complex number case can be easily reduced to the real case).
If $|\psi\rangle$ is a random unit vector in the same space of
$|\phi_A\rangle$ and $|\phi_B\rangle$, then
\begin{equation} \Prob\left[\sign(\langle \psi|\psi_A\rangle)\neq\sign(\langle
\psi|\psi_B\rangle)\right]\ =\ \theta/\pi.\label{eqn:angle}
\end{equation}
Hence, in order to estimate $\cos\theta$ with error term $\delta$,
it suffices to estimate $\theta/\pi$ to some error term $O(\delta)$
using the above equality checking of signs. Obviously this can
be done by a SMP protocol, and by a simple application of 
Chernoff Bound, requires $O\left(\ln\frac{1}{\eps}/\delta^2\right)$ repetitions. With
$\delta=O(\epsilon/C^2)$, this is $O\left(C^4\ln\frac{1}{\eps}/\epsilon^2\right)$ bits.

\comments{
A minor technical issue is that both $|\tilde\phi_A\rangle$
and $\tilde\phi_B\rangle$ may be complex vectors. But this can be
solved by separating their real and complex parts and apply the above
procedure on two pairs of the separated vectors, taking into 
account that $p$ is a real number.
}

We note that \cite{BaconT03} gives a procedure along the lines
of checking equality of signs but it produces
a random $\pm 1$ variable whose expectation is precisely
$\cos\theta$, though this is not asymptotically
advantageous. 

We summarize the above discussion as the basis for our future discussions.
\begin{theorem}[\cite{Kremer:1995:comm, GoemansW94}]\label{thm:estimate}
Suppose the acceptance probability of a quantum protocol can be 
expressed as $\langle \psi_A|\psi_B\rangle$, where
$|\psi_A\rangle$ and $|\psi_B\rangle$ can be prepared
by each party individually. Furthermore,  for some number $C$, 
$\||\psi_A\rangle\|\le C$, and $\||\psi_B\rangle\|\le C$. Then
there is a classical SMP protocol with shared coins that uses
$O\left(C^4{\ln\frac{1}{\eps}}/{\epsilon^2}\right)$ bits and 
whose acceptance probability deviates from that of the protocol
by at most $\epsilon$.
\end{theorem}

\section{The main theorem}
In this section, we formally define the classical
communication complexity and the diamond norm
of bipartite quantum operators, 
and derive an upper bound on the former in terms of the latter.
We shall focus
on the following case: that the measurement gives two outcomes,
and that the dimensions of the two systems are the same.
Our results can be extended trivially to more general cases.

We use script letters $\Space{N}$, $\Space{M}$, $\Space{F}$, $\cdots$,
to denote Hilbert spaces, and $\linear(\Space{N})$ to denote
the space of operators on $\Space{N}$. The identity operator
on $\Space{N}$ is denoted by $I_{\Space{N}}$, and the identity 
superoperator on 
$\linear(\Space{N})$ is denoted by ${\mathbf I}_{\Space{N}}$.
Recall that a {\em positive-operator-valued measurement (POVM)}
on a Hilbert space $\mathcal{H}$ 
is a set of positive semidefinite operators
$\{ Q_1, Q_2, \cdots, Q_m\}$ on $\mathcal{H}$,
such that $\sum_{i=1}^m Q_i=I_\mathcal{H}$.
Each $Q_i$ is called a {\em measurement element},
and corresponds to the measurement outcome $i$.
We may refer to a semidefinite operator $Q$, $0\le Q\le 1$,
as a {\em measurement element} of the implicit binary
POVM $\{ Q, I-Q\}$.
For more details on the foundations of quantum
information processing, refer to the textbook~\cite{NC00a}.

\begin{trivlist}
\item{\bf Classical simulation of quantum measurements.}
In this subsection we define the central concept of this paper:
the classical communication complexity of quantum measurements.

Let $Q$ be measurement element acting
on a bipartite system $AB$.
Let $|E\rangle_{A'B'}$ be a bipartite state, where $A'$ ($B'$) includes 
$A$ ($B$) as a subsystem.
Let $R_A$ and $R_B$ be physically realizable operators 
acting on system $A'$ and $B'$, respectively.
Denote by $\mu(Q, |E\rangle, R_A, R_B)$ the probability
\[ \mu(Q, |E\rangle, R_A, R_B)\defeq \trace(Q R_A\otimes R_B (|E\rangle\langle E|)).\]

\begin{definition}
Let $\eps\in [0, 1/2)$, and $Q$ be a measurement elements.
The {\em classical communication complexity}
of $Q$ with precision $\epsilon$,
denoted by $\ccc_\eps(Q)$, 
is the minimum number $k$ such that for any
$|E\rangle$, $R_A$ and $R_B$ described above,
there is a classical communication protocol between
two parties Alice and Bob that satisfies the following
conditions:
\begin{enumerate}
\item The input of Alice (Bob) is a classical description of
$|E\rangle$, and a classical description of $R_A$ ($R_B$);
\item The output is a random binary variable of which
the expected value $p$ satisfies
\[ |p-\mu(Q, |E\rangle, R_A, R_B)|\le \epsilon.\]
\item The protocol exchanges $\le k$ bits and 
is allowed to use an unlimited amount of shared randomness.
\end{enumerate}
\end{definition}

\item{\bf The diamond norm on bipartite operators.}
Let $\Space{N}$ be a Hilbert space and
$T: \linear(\Space{N})\to\linear(\Space{N})$ be a superoperator.
The {\em diamond norm} on super operators is defined as 
(c.f. \cite{Kitaev:2002:book})
\[ \|T\|_\diamond \defeq \inf\{ \|A\|\|B\| : 
\trace_{\mathcal{F}}(A\cdot B^\dagger)=T, \ 
A, \ B\in\linear(\Space{N}, \Space{N}\otimes\Space{F})\}.\]
For our application, the following alternative characterization
of the diamond norm is more convenient.
\begin{lemma}[e.g., \cite{Kitaev:2002:book}]\label{lm:diamondEquiv}
For any superoperator $T$,
\[\|T\|_\diamond=\min\ \{\ 
\sqrt{\|\sum_t A_t^\dagger A_t\|}\cdot\sqrt{\|\sum_t B_t^\dagger B_t\|}\ : \ 
\ 
A_t, \ B_t\in\linear(\Space{N}),\ T=\sum_t A_t\cdot B_t^\dagger
\ \}.\]
\end{lemma}

Let $\Space{N}_A$, $\Space{N}_B$, and $\Space{N}$
be Hilbert spaces of the same dimension.
We fix an isomorphism between any two of them.
For an operator in one space, 
we use the same notation for its images and preimages,
under the isomorphisms, in the other spaces.

Let $Q\in\linear(\Space{N}_A\otimes\Space{N}_B)$ be 
a bipartite operator and $Q=\sum_{t} A_t\otimes B_t^\dagger$,
for some $A_t\in\linear(\Space{N}_A)$, and
$B_t\in\linear(\Space{N}_B)$. Define a mapping $\mathcal{T}$ from bipartite
operators on $\Space{N}_A\otimes\Space{N}_B$ to superoperators
$\linear(\Space{N})\to\linear(\Space{N})$
by mapping $Q\mapsto \mathcal{T}(Q)\defeq \sum_{t} A_t\cdot B_t^\dagger$.
It can be easily verified that the mapping is independent
of the choice of the decomposition of $Q$ and is indeed
an isomorphism.

\begin{definition}
Let $Q\in\linear(\Space{N}_A\otimes\Space{N}_B)$ be
a bipartite operator. The {\em diamond norm} of $Q$, denoted
by $\|Q\|_\diamond$, is $\|Q\|_\diamond\defeq\|\mathcal{T}(Q)\|_\diamond$.
\end{definition}

By Lemma~\ref{lm:diamondEquiv}, for any $Q$,
\[
\|Q\|_\diamond = \min\{ 
\sqrt{\|\sum_t A_t^\dagger A_t\|}\cdot\sqrt{\|\sum_t B_t^\dagger B_t\|}\ :\ 
A_t\in\linear(\Space{N}_A), \ B_t\in\linear(\Space{N}_B),\ Q=\sum_t A_t\otimes B_t^\dagger
\ \}.\]

Note that if a superoperator $T=A\cdot B$ for some $A, B\in \linear(\Space{N})$,
$\|T\|_\diamond=\|A\|\cdot\|B\|$. Therefore the diamond norm on 
bipartite operators is a tensor norm:
\begin{lemma}\label{lm:tensor}
If $K=A\otimes B$, $\|K\|_\diamond=\|A\|\cdot\|B\|$.
\end{lemma}

A nice property of the superoperator diamond norm is
that it is ``stable'', i.e., it remains unchanged when
tensored with the identity operator on an additional space \cite{Kitaev:2002:book}. 

\begin{lemma}\label{res:diamondStable}
Let $\Space{N}$, $\Space{M}$,
and $\Space{F}$ be Hilbert spaces, and
$T: \linear(\Space{N})\to\linear(\Space{M})$ be a superoperator.
Then $\|{\mathbf I}_{\Space{F}}\otimes T\|_\diamond=
\|T\|_\diamond$.
\end{lemma}

This stability property carries over to our diamond norm
and is important for our applications.
Let $\Space{F}_A$ and $\Space{F}_B$
be Hilbert spaces of the same dimension, 
and $Q\in\linear(\Space{N}_A \otimes\Space{N}_B)$.
Denote by $Q_{\Space{F}_A, \Space{F}_B}$ the bipartite operator
$Q\otimes I_{\Space{F}_A\otimes\Space{F}_B}$, where
the two subsystems are $\Space{N}_A\otimes\Space{F}_A$
and $\Space{N}_B\otimes\Space{F}_B$.
\begin{lemma}\label{lm:stable}
For any $Q$,
$\|Q_{\Space{F}_A, \Space{F}_B}\|_\diamond=\|Q\|_\diamond$.
\end{lemma}

If $Q$ is a measurement element of a POVM acting
on a Hilbert space
of dimension $K$, a trivial upper bound on $\|Q\|_\diamond$ is
$K^2$, as each entry of the matrix of $Q$ under any
orthonormal basis has a modulus bounded by $1$.

\begin{proposition}\label{res:trivial}
If a bipartite operator $Q$ is measurement element of a POVM acting
on a Hilbert space of dimension $K$, then $\|Q\|_\diamond\le K^2$.
\end{proposition}

This bound is not far from being optimal for $\textsc{IP}_n$,
in which case $K=2^{2n}$.
To prove a lower bound on $\|\textsc{IP}_n\|_\diamond$, we
use a remarkable dual characterization of the diamond norm
(e.g., \cite{Kitaev:2002:book}, Theorem 11.1).
Let $T: L(\mathcal{N})\rightarrow L(\mathcal{N})$ be superoperator
and $\mathcal{G}$ be a space of the same dimension as $\mathcal{N}$.
Then
\begin{equation}\label{eqn:dualdiamond}
\|T\|_\diamond = \sup_{\rho\in L(\mathcal{N}\otimes\mathcal{G}),
\rho\neq 0} \frac{ \|(T\otimes {\mathbf I}_{\mathcal{G}})(\rho)\|_{\tr}}{\|\rho\|_{\tr}}.
\end{equation}

\begin{proposition}\label{res:ipn}
For the $\textsc{IP}_n$ operator defined in Equation~\ref{eq:ipn},
$\|\textsc{IP}_n\|_\diamond\|\ge 2^{n-1}$.
\end{proposition}

\begin{proof}
By definition, 
\[ \mathcal{T}(\textsc{IP}_n)=\sum_{x, y\in\{0, 1\}^n, x\cdot y=1}
\ |x\rangle\langle x|\cdot |y\rangle\langle y|.\]

We set $\rho=\sum_{x, y} |x\rangle\langle y| \otimes I_{\Space{G}}$
in Equation~\ref{eqn:dualdiamond}, resulting in
\[ \|\mathcal{T}(\textsc{IP}_n)\|_\diamond \ \ge\  \frac{1}{2^{n}} 
\ \left\|\sum_{x, y\in\{0, 1\}^n, x\cdot y=1} |x\rangle\langle y|
\right\|_{\tr}.\]

The right-hand-side is at least 
\[ \frac{1}{2^n} \ \textrm{trace}\  
\left(\sum_{x, y\in\{0, 1\}^n, x\cdot y=1} |x\rangle\langle y|
\right)
\ \ge\  2^{n-1}.\]

Thus $\|\textsc{IP}_n\|_\diamond\ge 2^{n-1}$.
\end{proof}

We conclude this subsection by noting that
our diamond norm on bipartite operators appears natural
in connection with the following matrix analogy of the Cauchy
Schwartz Inequality.
\begin{theorem}[Joci{\'c}~\cite{Jocic:1999}] For any operators $A_t$ and $B_t$,
\begin{equation}\label{eqn:matrixCauchy}
\|\sum_t A_t\otimes B^\dagger_t\|\le \sqrt{\|\sum_t A_t^\dagger A_t\|}
\cdot\sqrt{\|\sum_t B^\dagger_tB_t\|}.
\end{equation}
\end{theorem}
Hence, if $\|Q\|_\diamond$ is precisely the smallest
right-hand-side when $A_t$ and $B_t$ are such that
$Q=\sum_t A_t\otimes B_t^\dagger$. Inequality~(\ref{eqn:matrixCauchy})
may actually be proved by the same approach that we use
to prove Theorem~\ref{thm:diamond} below.

\item{\bf Upper bounding $\ccc(Q)$ by the diamond norm.}
We now use the diamond norm to derive an upper bound 
on $\ccc_\eps(Q)$.
Recall that if $\Space{M}$ and $\Space{N}$ are
two Hilbert spaces, an {\em isometric embedding}
$U: \Space{M}\rightarrow \Space{N}$ is a linear map
that satisfies $U^\dagger U=I_{\Space{M}}$.

\begin{theorem}\label{thm:diamond}
\newcommand{\sd}{\mathbf{SD}}
For any bipartite positive semidefinite operator $Q$ acting
on a Hilbert space of dimension $K$,
\begin{equation}\label{eqn:diamond}
\ccc_\eps(Q)=O\left(\|Q\|_\diamond^2\cdot {\ln\frac{1}{\eps}}/{\eps^2}\right).
\end{equation}
In particular $\ccc_\eps(Q)=O(K^4\log\frac{1}{\epsilon}/\epsilon^2)$.
Furthermore, the upper bound (\ref{eqn:diamond}) can be achieved by a SMP protocol
with shared randomness.
\end{theorem}
\begin{proof} 
Without loss of generality, assume that on receiving
their portions of $|E\rangle$, Alice and Bob apply an isometric
embedding $U: \Space{M}_A\to\Space{N}_A\otimes\Space{F}_A$,
and $V: \Space{M}_B\to\Space{N}_B\otimes\Space{F}_B$, respectively,
for some Hilbert spaces $\Space{F}_A$ and $\Space{F}_B$ with an
equal dimension. 
The distribution resulted from
Charlie's measuring $Q$ on $\Tr_{\Space{F}_A, \Space{F}_B}
\left(( U\otimes V)|E\rangle\langle E| (U\otimes V)^\dagger\right)$
is the same as that of Charlie applying
$Q_{\Space{F}_A, \Space{F}_B}$
on the larger state $(U\otimes V)|E\rangle\langle E|(U\otimes V)^\dagger$.
\comments{
When $Q'$ is regarded as a bipartite operator with
$\Space{F}_A$ and $\Space{F}_B$ separated,
$\mathcal{T}(Q')= {\mathbf I}_{\Space{F}}\otimes \mathcal{T}(Q)$,
where $\Space{F}$ is a Hilbert space of the same dimension
as $\Space{F}_A$ and $\Space{F}_B$, and ${\mathbf I}_{\Space{F}}$
is the identity superoperator on $\linear(\Space{F})$.
}
By Lemma~\ref{lm:stable}, $\|Q_{\Space{F}_A, \Space{F}_B}\|_\diamond
=\|Q\|_\diamond$. Therefore, to prove the theorem we need
only to consider isometric embeddings $U:\Space{M}_A\to\Space{N}_A$ and 
$V:\Space{M}_A\to\Space{N}_B$.

Without loss of generality, we assume that Alice and Bob
have agreed on a Schmidt decomposition
$|E\rangle=\sum_i \sqrt{p_i} |i\rangle_A\otimes |i\rangle_B$,
for some $p_i\ge 0$, $\sum_i p_i=1$, and for an orthonormal
basis $\{ |i\rangle \}$. Denote by $|i_A\rangle\defeq U|i\rangle$,
and $|i_B\rangle\defeq V|i\rangle$.
Then the message that Charlie
receives is $|\bar E\rangle\defeq(U\otimes V)|E\rangle=\sum_i \sqrt{p_i}
|i_A\rangle\otimes|i_B\rangle$.  

Suppose $\|Q\|_\diamond$ is achieved under the 
decomposition $Q=\sum_t A_t\otimes B_t^\dagger$,
with which if 
$Q_A\defeq\sum_t A_t^\dagger A_t$, and,
$Q_B\defeq\sum_t B_t^\dagger B_t$,
we have $\|Q_A\|=\|Q_B\|=\superNorm{Q}$.
With those definitions, we have
\[p=\langle \bar E |Q|\bar E\rangle =
\sum_{i, j, t} \sqrt{p_ip_j}\ \langle i_A|A_t|j_A\rangle\cdot
\langle i_B|B^\dagger_t|j_B\rangle.\]

Define two vectors
\begin{equation}\label{eqn:psiGA}
|\psi_A\rangle = \sum_{i, j, t} \sqrt{p_j}\
\langle j_A|A_t^\dagger|i_A\rangle \ |i, j, t\rangle,
\quad\textrm{and,}
\end{equation}
\begin{equation}\label{eqn:psiGB}
|\psi_B\rangle = \sum_{i, j, t} \sqrt{p_i}\
\langle i_B|B_t^\dagger|j_B\rangle \ |i, j, t\rangle.
\end{equation}
Then $p=\langle \psi_A|\psi_B\rangle$.
Further, with $\rho_A\defeq\sum_j p_j|j_A\rangle\langle j_A|$,
\[\langle \psi_A|\psi_A\rangle
=\sum_{i, j, t}
p_j |\langle j_A|A_t^\dagger|i_A\rangle|^2
=\trace(\rho_A Q_A)
\le \|Q_A\|={\superNorm{Q}}.\]

Similarly, $\langle\psi_B|\psi_B\rangle\le \|Q_B\|=\|Q\|_\diamond$. 
Therefore, by Theorem~\ref{thm:estimate},
the measurement scenario can be approximated by a 
classical SMP with shared coins to be
within an $\eps$ precision using 
$O\left(\|Q\|_\diamond^2{\ln\frac{1}{\eps}}/{\eps^2}\right)$ bits. 
This bound is $O(K^4\log\frac{1}{\epsilon}/\epsilon^2)$ as
$\|Q\|_\diamond=O(K^2)$ by Proposition~\ref{res:trivial}.
\end{proof}

\begin{remark}\label{rmk:alpha} One may improve the above upper bound
on $\ccc_\eps(Q)$ by a more carefully chosen $|\psi_A\rangle$
and $|\psi_B\rangle$ in Equation \ref{eqn:psiGA} and \ref{eqn:psiGB}.
More specifically,
let $\alpha\in[0, 1]$, define
\[|\psi^\alpha_A\rangle = \sum_{i, j, t} \sqrt{p_i^{\alpha} p_j^{1-\alpha}}\ 
\langle j_A|A_t^\dagger|i_A\rangle \ |i, j, t\rangle,
\quad\textrm{and,}
\]\[
|\psi^\alpha_B\rangle = \sum_{i, j, t} \sqrt{p_i^{1-\alpha} p_j^{\alpha}}\ 
\langle i_B|B_t^\dagger|j_B\rangle \ |i, j, t\rangle.
\]
One can verify that minimizing $\||\psi_A\rangle\|\cdot\||\psi_B\rangle\|$
over all decompositions of $Q$ gives rise to a tensor norm, which we do not
know if is stable under tensoring with identity superoperators.
Although we have 
not found any useful application of an $\alpha\ne 0$, we cannot rule out
the possibility that a carefully chosen $\alpha$ may give a better bound.
\end{remark}

\begin{remark}\label{rmk:otimes}
In the case that $|E\rangle$ is not entangled,
the same approach in Theorem~\ref{thm:diamond} can be used
to derive a systematic classical simulation. More specifically,
in this context we would like to estimate 
$p=\langle \phi_A\otimes\phi_B| Q |\phi_A\otimes\phi_B\rangle$,
for a state $|\phi_A\rangle$ known to Alice only and a state
$|\phi_B\rangle$ known to Bob only. For a decomposition 
of $Q=\sum_t A_t\otimes B_t^\dagger$, we define
\[|\psi_A\rangle = \sum_{t} \langle \phi_A|A^\dagger_t|\phi_A\rangle |t\rangle,
\quad \textrm{and,}\quad
 |\psi_B\rangle =
 \sum_{t} \langle \phi_B|B^\dagger_t|\phi_B\rangle  |t\rangle.
\]
Then $p=\langle \psi_A|\psi_B\rangle$. It can be verified that
\[\|Q\|_\otimes\defeq \inf\{ \|\psi_A\|\cdot\|\psi_B\| : Q=\sum_t A_t\otimes 
B_t^\dagger\}\]
defines a tensor norm and $\|Q\|_\otimes\le \|Q\|_\diamond$.
This approach gives a constant cost simulation of the elegant
quantum fingerprint protocol of Buhrman, Cleve, Watrous, and de Wolf \cite{BCWdW01a}
for testing equality of two input strings.
\end{remark}

\end{trivlist}
\section{Applications}
We now apply the above to derive classical upper bounds
on quantum communication complexity.
\begin{trivlist}
\item {\bf Quantum SMP with shared entanglement.}
If the quantum protocol is in the SMP model with
shared entanglement, we immediately have,
\begin{corollary}[of Theorem~\ref{thm:diamond}
]
\label{co:smp}
If in a quantum SMP protocol, Charlie applies the measurement
$P$, then the protocol can 
be simulated by a classical SMP
protocol with shared coins and using $O(\|P\|_\diamond^2)$ bits.
\end{corollary}

\item {\bf Twoway interactive quantum communication with shared
entanglement.}
Now consider the general twoway interactive quantum communication.
We need the following lemma due to Yao \cite{Yao:1993:circuit},
and the following formulation is from \cite{Razborov:2002:disj}:
\begin{lemma}[\cite{Yao:1993:circuit, Razborov:2002:disj}] \label{lm:Yao} 
Let $\mathcal{P}$ be a two-party interactive quantum
communication protocol that uses $q$ qubits.
Let $\Space{H}_A$ and $\Space{H}_B$ be the state spaces of Alice and Bob,
respectively. 
For an input $(x, y)$, denote by
$|\Phi_{x, y}\rangle_{AB}$ the joint state of Alice, Bob 
before the protocol starts.
Then there exist linear operators $A_h\in \linear(\Space{H}_A)$,
and $B_h\in\linear(\Space{H}_B)$, for each $h\in\{0, 1\}^{q-1}$,
such that 
\begin{enumerate}
\item[(a)] $\|A_h\|\le 1$ and $\|B_h\|\le 1$ for all $h\in\{0, 1\}^{q-1}$;
\item[(b)] the acceptance probability of $\mathcal{P}$
on input $x$ and $y$ is $\|P|\Phi_{x, y}\rangle\|^2$,
where $P\defeq \sum_{h\in\{0, 1\}^{q-1}} A_h\otimes B_h.$	
\end{enumerate}
\end{lemma}
We are now ready to prove Theorem~\ref{thm:twoway}.
\begin{proofof}{Theorem~\ref{thm:twoway}}
Let $|E\rangle_{AB}$ be the shared entanglement,
For an $n$-bit binary string $x$,
denote by $U_x$ the isometric embedding from $\mathbb{C}$ to $\mathbb{C}^{\otimes 2^n}$
that maps $c\mapsto c |x\rangle$. 
Let $P$, $A_h$, and $B_h$ be those in Lemma~\ref{lm:Yao}.
Then the quantum protocol gives rise to a measurement scenario
in which the measurement is $P^\dagger P$, the shared entanglement
is $|E\rangle$,
and on an input pair $(x, y)$, Alice's private operator is 
$U_x$ and that of Bob is $U_y$.

By Theorem~\ref{thm:diamond},
the acceptance probability can be estimated with $O(\|P^\dagger P\|^2_\diamond)$ bits of communication
in the SMP model with shared randomness.
Since $\|\cdot\|_\diamond$
is a tensor norm, we have
\[\|P^\dagger P\|_\diamond \le 
\sum_{h, h'}\ \| \left((A_{h'})^\dagger A_h\right)
\otimes \left((B_{h'})^\dagger B_h\right)\|_\diamond
=\sum_{h, h'} \|A_h\|\|A_{h'}\|
\|B_h\|\|B_{h'}\|\le 2^{2(q-1)}.\]

The last inequality is because
$\|A_h\|\le 1$ and $\|B_h\|\le 1$
for all $h$.
Hence the acceptance probability can be estimated 
by a classical SMP protocol using $\exp(O(q))$ bits.
\end{proofof}

Corollary \ref{co:constLog} follows trivially from the above
by setting $q$ to be a constant. Corollary \ref{res:q2qsmp}
follows immediately from Theorem~\ref{thm:YaoFinger}
and Corollary \ref{co:constLog} together with the following observation.
\begin{lemma}\label{res:consteq}
If a communication complexity problem has a classical twoway
protocol with shared randomness and $b$ bits of cost, it
has a classical SMP protocol with shared randomness and
$O(b2^{b/2})$ bits of communication.
\end{lemma}
\begin{proof}
Fix a twoway protocol for the problem in which
Alice sends $b_A$ bits and bob sends $b_B$ bits.
To simulate this protocol in the SMP model with shared
randomness, Alice sends the referee $2^{b_B}$ strings
each of which has $b_A$ bits and is consistent
with her input and a string of $b_B$ bits interpreted
as Bob's messages. Bob applies the same strategy
to sends $2^{b_A}$ strings of $b_B$ bits.
The referee is then able to reconstruct a string
of $b$ bits, which is precisely the transcript of
communication in the original protocol with the
same input and random string. Hence by outputting
the last bit of the reconstructed message, this SMP
protocol achieves the same error probability of the original
protocol. The cost of the simulating protocol is
$2^{b_A}b_B+2^{b_B}b_A=O(b2^{b/2})$ bits.
\end{proof}

\item {\bf Simulating quantum correlations.}
We shall define precisely what we mean by simulating
quantum correlations.

We define a {\em quantum measurement game}
as a triple $G=( |E\rangle_{AB}, \mathcal{P}_A,
\mathcal{P}_B )$, where
$|E\rangle_{AB}$ is a bipartite quantum
state, $\mathcal{P}_A$, $\mathcal{P}_B$
are sets of possible measurements on the system $A$
and the system $B$, respectively. Let $\mathcal{V}_A$ 
($\mathcal{V}_B$, respectively) be the set of possible
measurement outcomes of $\mathcal{P}_A$ ($\mathcal{P}_B$,
respectively). For $P_A\in\mathcal{P}_A$
and $P_B\in\mathcal{P}_B$, denote by $\omega_G(P_A, P_B)$
the distribution of the measurement outcomes when
$P_A\otimes P_B$ is applied to $|E\rangle$.

A {\em classical simulation} of a quantum measurement
game $G=(|E\rangle_{AB}, \mathcal{P}_A, \mathcal{P}_B)$
is a classical communication protocol between
two parties Alice and Bob, who
start with an unlimited mount of shared
randomness, and Alice
has the classical description of an element $P_A\in\mathcal{P}_A$,
while Bob has the classical description of an element $P_B\in\mathcal{P}_B$.
At the end of the protocol, Alice (and Bob)
outputs an element from $\mathcal{V}_A$ (
$\mathcal{P}_B$, respectively),
resulting in a distribution $\tilde\omega(P_A, P_B)$.

We are now able to rigorously state Theorem~\ref{thm:bell}.
Recall that the statistical distance between two distributions
$\pi=(p_1, \cdots, p_n)$ and $\tilde\pi=(\tilde p_1, \cdots, \tilde p_n)$ is 
$\|\pi -\tilde\pi\|_1\defeq\sum_{i}|p_i-\tilde p_i|$.

\begin{theorem}\label{thm:bellRig}
Let $G=(|E\rangle_{AB}, \mathcal{P}_A, \mathcal{P}_B)$
be a quantum measurement game, $m=|\mathcal{V}_A|\cdot|\mathcal{V}_B|$,
and $\epsilon\in\mathbb{R}$, $0\le \epsilon <1$.
There is a classical simulation of $G$ that exchanges
$O(\frac{m^3}{\epsilon^2}\cdot\ln \frac{m}{\epsilon})$ number of bits
and the output distribution $\tilde\omega(P_A, P_B)$ for 
any $P_A\in\mathcal{P}_A$ and $P_B\in\mathcal{P}_B$ satisfies
\[ \| \tilde\omega(P_A, P_B) - \omega_G(P_A, P_B)\|_1\le \epsilon.\]
In particular, the simulation cost is $O(\log\frac{1}{\epsilon}/\epsilon^2)$ 
if $m=O(1)$. 
\end{theorem}

\begin{proof}
Recall that a POVM measurement can be expressed as a physically realizable
operator followed by a projective measurement (see, e.g., \cite{Kitaev:2002:book}).
Thus we can assume without loss of generality that there exist
projections $P_A^v$, $v\in\mathcal{V}_A$, and $P_B^{v'}$, $v'\in\mathcal{V}_B$,
such that for each $P_A\in\mathcal{P}_A$ ($P_B\in\mathcal{P}_B$),
there is an isometric embedding $U_A$ ($U_B$) so that $P_A$  ($P_B$) consists
of the measurement elements $\{U_A^\dagger P^v U_A : v\in \mathcal{V}_A\}$
($\{U_B^\dagger P^{v'} U_B : v'\in\mathcal{V}_B\}$).

Fix a pair of measurements $(P_A, P_B)$.
In the classical simulation protocol, Alice and Bob first
compute the probability of outputting $(v, v')$
to be within $\epsilon/m$ error, for each $v\in\mathcal{V}_A$
and $v'\in\mathcal{V}_B$.
They then output $(v, v')$ according to the probabilities
computed. Thus $\tilde\omega(P_A, P_B)$ is within $\epsilon$ statistical
distance to $\omega(P_A, P_B)$.

Fix a pair of possible outcome $(v, v')$. 
Let $P^{v, v'}\defeq P_A^v\otimes P_B^{v'}$.
Then by Lemma~\ref{lm:tensor},
$\|P^{v, v'}\|_\diamond=\|P^v_A\|\cdot\|P^{v'}_A\|\le 1$.
The estimation of $\omega_G(P_A, P_B)$ now becomes
the simulation of the measurement element $P^{v, v'}$
with the initial state being $|E\rangle$, and the local
physically realizable operators being $U^\dagger_A\cdot U_A$
and $U_B^\dagger \cdot U_B$.

Hence by Theorem~\ref{thm:diamond}, the probability of
observing outcome $(v, v')$ can be calculated to be within
$O(\eps/m)$ precision by
by a classical protocol using 
$O\left({m^2\ln(m/\eps)}/{\eps^2}\right)$ bits.
Thus the overall simulation cost is
$O\left({m^3\ln(m/\eps)}/{\eps^2}\right)$ bits, 
which is $O(\log\frac{1}{\epsilon}/\epsilon^2)$ when
$m=O(1)$.
\end{proof}

\end{trivlist}
\section{Conclusion and open problems}
\newcounter{openProblem}
\setcounter{openProblem}{0}
\newcommand{\newP}{\addtocounter{openProblem}{1}\item \mbox{\textsc{Problem \arabic{openProblem}.}}{\ }}
A central mission of quantum information theory is to understand
quantitatively the boundaries between quantum and classical processes.
In this paper, we provide an alternative measure of nonlocalness
of bipartite quantum measurements: the minimum amount of classical
communication required to simulate the quantum measurement.
After defining this concept, we give an upper bound by constructing
a simulating protocol. The upper bound is expressed in terms of
a tensor norm, which captures nonlocalness in its own way, and may be
of independent interest and further applications. Variants of our protocol
also lead to variants of the main upper bound in terms of other tensor norms.

We then apply our upper bound to the classical simulation
of quantum communication protocols and the construction of 
local hidden variable models augmented with classical communication.
In particular, we show that quantum and classical communication protocols
with unlimited shared entanglement or randomness
compute the same set of functions, if the amount of communication is constant.
We also show that local measurements of an entangled state can be simulated
by a local hidden variable model with a constant amount of communication,
as long as the number of measurement outcomes is constant.

Our study is only the first step toward understanding
the classical communication complexity of bipartite measurements.
An obvious open problem is to prove or disprove
that the bound in Theorem~\ref{thm:diamond} is tight.
Another basic question is to prove a strong lower bound
(exponential in the number of qubits) on $\ccc(Q)$ for some $Q$.

It would be interesting to relate $\ccc(Q)$ to
other measures of nonlocality, such as the entanglement capacity,
and the minimum number of elementary gates,
or the amount of time for evolving some elementary Hamiltonian,
needed to approximate $Q$. It is conceivable that by the comparisons
of those measures may lead to a unique and representative measure of nonlocalness.
 
A recent progress on the question of the usefulness of
quantum entanglement was made by Gavinsky~\cite{Gavinsky:2006}, in
which he showed that entanglement is responsible for 
exponential savings for some communication tasks and
in some restricted models. Whether or not entanglement
could result in exponential savings for the more standard two-way
communication model and for the computation of functions
remains unsolved.
Can our result on removing the entanglement be strengthened
to that one can always use an amount of entanglement 
linear in size of the messages, with at most a logarithmic additive
term?

The cost of our protocol for simulating quantum correlations
depends linearly on the number of measurement outcomes. Is this
dependence necessary or can one dramatically reduce it?

Finally, it appears a very promising direction to us
to further exploring the connections of tensor norms and 
nonlocalness of quantum states and operations. 

\section{Acknowledgments}
We are indebted to Wei Huang, Amnon Ta-Shma, and the anonymous reviewers 
for their valuable suggestions on improving the
presentation of this paper.
%
\bibliographystyle{abbrv}

\end{document}